# On the structure of symmetric self-dual Lie algebras

José M. Figueroa-O'Farrill[†]

*Department of Physics, Queen Mary and Westfield College
Mile End Road, London E1 4NS, UK*

and

Sonia Stanciu[‡]

*ICTP, P.O. Box 586, I-34100 Trieste, ITALY*

ABSTRACT

A finite-dimensional Lie algebra is called (symmetric) self-dual, if it possesses an invariant nondegenerate (symmetric) bilinear form. Symmetric self-dual Lie algebras have been studied by Medina and Revoy, who have proven a very useful theorem about their structure. In this paper we prove a refinement of their theorem which has wide applicability in Conformal Field Theory, where symmetric self-dual Lie algebras start to play an important role due to the fact that they are precisely the Lie algebras which admit a Sugawara construction. We also prove a few corollaries which are important in Conformal Field Theory.

[†] e-mail: j.m.figueroa@qmw.ac.uk

[‡] e-mail: sonia@ictp.trieste.it

§1 INTRODUCTION AND MOTIVATION

For most physical applications, reductive Lie algebras are the most natural Lie algebras to consider. This is because they are the Lie algebras of the compact Lie groups, which have played a privileged role in physical theories. Reductive Lie algebras are completely classified, since they are direct products of abelian and semisimple Lie algebras, and essentially everything is known about them and their representations, at least the finite-dimensional ones. However by any reasonable measure, reductive Lie algebras are rare; and comparatively little is known about their nonreductive counterparts. The Levi-Malcev theorem reduces the classification problem for general Lie algebras to that of semi-direct products (*i.e.*, split extensions) of semisimple and solvable Lie algebras; but already classifying solvable Lie algebras seems to be as hard as classifying Lie algebras in general: by brute force one can classify all Lie algebras of dimension $\leq 5$, and restricting oneself to solvable Lie algebras does not get one any further (although all nilpotent six-dimensional Lie algebras are known). Therefore in order to probe the space of Lie algebras one could hope to benefit by restricting oneself to a class of Lie algebras including the reductive Lie algebras but which are still special enough to allow for a classification. One property shared by all reductive Lie algebras is the existence of an invariant metric; that is, an invariant nondegenerate symmetric bilinear form. We will call Lie algebras possessing an invariant metric *symmetric self-dual* Lie algebras, and they comprise a nontrivial generalisation of reductive Lie algebras. Although no classification exists to this date, there exists a structure theorem [1] which tells us in principle how to construct Lie algebras with an invariant metric starting from the reductive Lie algebras. In fact, reductive Lie algebras can be obtained from simple Lie algebras and the one-dimensional Lie algebra by the operation of direct sum. What Medina and Revoy found in [1] is that all Lie algebras with an invariant metric can be obtained from the same ingredients provided that we introduce a new operation—the *double extension*—which generalises the semidirect product in a nontrivial way.

The importance of symmetric self-dual Lie algebras in Conformal Field Theory (and via CFT in String Theory) has to do with the following curious fact [2] [3]: symmetric self-dual Lie algebras are precisely the Lie algebras for which a Sugawara construction exists. This fact may not appear so surprising if one assumes that the relation between the Sugawara construction and the WZW model persists in the nonreductive case: a WZW model needs for its definition a Lie group possessing a bi-invariant metric and this condition translates, at the level of the Lie algebra, into the statement that its Lie algebra should possess an invariant metric. Interestingly enough, the relation between the WZW model and the Sugawara construction does persist in the nonreductive case [4], but the proof of this statement is not immediate and happens to necessitate detailed knowledge of the structure of symmetric self-dual Lie algebras, and in particular some refinements of the structure theorem in [**1**]. The purpose of this letter is to collect those results on the structure of symmetric self-dual Lie algebras that were used in [**4**]. Of necessity, those results are of a less physical nature than their applications, and we felt it inappropriate to include them together; hence the present paper.

This letter is organised as follows. In Section 2 we assemble some basic properties of symmetric self-dual Lie algebras and some properties of their ideals, which will be needed when we review the structure theorem of Medina and Revoy in Section 3. In Section 4 we define the double extension of a symmetric self-dual Lie algebra by a second Lie algebra and we work out some explicit formulas which we will need later. In Section 5 and Section 6 we prove some useful refinements and corollaries of the structure theorem. In Section 7 we comment briefly on the applications of this formalism to Conformal Field Theory and String Theory; and finally in Section 8 we mention some possible extensions and open problems. The paper also includes an appendix of a result on splittings of exact sequences involving Lie algebras. This result is used by Medina and Revoy, but we have not found a reference for it anywhere and we were forced to rederive it ourselves. We include it here for completeness.

§2   BASIC PROPERTIES OF SYMMETRIC SELF-DUAL LIE ALGEBRAS

In this section we set up the notation and we introduce the necessary concepts about symmetric self-dual Lie algebras that we will need in the sequel.

DEFINITION 2.1. Let $\mathcal{C}$ denote the class of pairs $(\mathfrak{g}, \langle -, - \rangle)$, where $\mathfrak{g}$ is a finite-dimensional Lie algebra and $\langle -, - \rangle$ is a nondegenerate ad-invariant symmetric bilinear form on $\mathfrak{g}$. We shall call such a bilinear form simply an *invariant metric*, and we shall (tentatively) call an element of $\mathcal{C}$ a *symmetric self-dual Lie algebra*.

REMARK 2.2. We should hasten to add that the nomenclature is by no means standard. French authors call these Lie algebras "orthogonal," whereas others calls them "self-dual." The name "self-dual" presumably comes from the fact that the adjoint representation is equivalent to the coadjoint representation. But clearly for this to be the case, all that one requires is a nondegenerate invariant bilinear form on $\mathfrak{g}$, but not one that need be symmetric. After consulting with Gregg Zuckerman, who seems to have inspired "self-dual" in [**5**], we have chosen the compromise "symmetric self-dual," since it causes no confusion with the Lie algebras of the orthogonal Lie groups and does not preempt the term self-dual for their more general cousins. Nevertheless, since only symmetric self-dual Lie algebras will play a role in this paper, *we will use the term "self-dual" from now on to mean "symmetric self-dual" unless otherwise stated.*

Let us first mention some minor matters of notation. If a Lie algebra $\mathfrak{g}$ should decompose as a direct sum of *subspaces A and B* we will write $\mathfrak{g} = A \oplus B$. If moreover, the subspaces are *ideals*, so that the decomposition is one of Lie algebras, then we will write $\mathfrak{g} = A \times B$. It will prove convenient to introduce some nomenclature for particular subspaces of a self-dual Lie algebra depending on how the metric behaves on them.

DEFINITION 2.3. Let $(\mathfrak{g}, \langle -, - \rangle)$ be a self-dual Lie algebra. For any subspace $V \subset \mathfrak{g}$, let $V^\perp = \{w \in \mathfrak{g} \mid \langle w, v \rangle = 0 \text{ for all } v \in V\}$. Notice that the operation $V \mapsto V^\perp$ is involutive, so that $(V^\perp)^\perp = V$. We say that

$$\text{V is } \textit{isotropic} \Leftrightarrow V \subset V^\perp ,$$
$$\text{V is } \textit{coisotropic} \Leftrightarrow V \supset V^\perp ,$$
$$\text{V is } \textit{lagrangian} \Leftrightarrow V = V^\perp ,$$
$$\text{V is } \textit{degenerate} \Leftrightarrow V \cap V^\perp \neq 0 ,$$
$$\text{V is } \textit{nondegenerate} \Leftrightarrow V \cap V^\perp = 0 .$$

Let $(\mathfrak{g}, \langle -, - \rangle)$ be a self-dual Lie algebra. We define the centralizer $Z_\mathfrak{g}(V)$ of a subspace $V \subset \mathfrak{g}$ as all those elements in the Lie algebra which commute with all elements of the subspace; that is $Z_\mathfrak{g}(V) \equiv \{w \in \mathfrak{g} \mid [w, v] = 0 \text{ for all } v \in V\}$. For a self-dual Lie algebra, centralizers and ideals are intimately linked, as the following lemma suggests:

LEMMA 2.4. $I \subset \mathfrak{g}$ is an ideal if and only if $I^\perp \subset Z_\mathfrak{g}(I)$.

PROOF: This follows immediately by the invariance of the metric. Indeed, $\langle [\mathfrak{g}, I], I^\perp \rangle = \langle \mathfrak{g}, [I, I^\perp] \rangle$, from where it follows that $[\mathfrak{g}, I] \subset I^{\perp\perp} = I$ if and only if $[I, I^\perp] = 0$. □

The center of a self-dual Lie algebra can be also characterised very easily. In fact,

LEMMA 2.5. *Let $[\mathfrak{g}, \mathfrak{g}]$ denote the first derived ideal and $Z(\mathfrak{g})$ be the center. Then $[\mathfrak{g}, \mathfrak{g}]^\perp = Z(\mathfrak{g})$.*

PROOF: $x \in Z(\mathfrak{g}) \Leftrightarrow [x, y] = 0 \; \forall y \Leftrightarrow \langle [x, y], z \rangle = 0 \; \forall y, z \Leftrightarrow \langle x, [y, z] \rangle = 0 \; \forall y, z \Leftrightarrow x \in [\mathfrak{g}, \mathfrak{g}]^\perp$. □

First of all notice that if $I \subset \mathfrak{g}$ is an ideal, so is $I^\perp$. Recall that an ideal $I \subset \mathfrak{g}$ is minimal if it does does not properly contain another nontrivial ideal $J \subset \mathfrak{g}$. In other words, if $I \subset \mathfrak{g}$ is a minimal ideal and if $J \subset \mathfrak{g}$ is another ideal with $J \subset I$, then either $J = 0$ or $J = I$. Below we list some properties of minimal ideals that we shall need in the proof of the structure theorem or its refinements.

PROPOSITION 2.6. *Let $(\mathfrak{g}, \langle-,-\rangle)$ be self-dual, and let $I \subset \mathfrak{g}$ be a minimal ideal. Then,*

(1) *If $I$ is nondegenerate, then it is a factor, and hence simple or one-dimensional;*

(2) *If $I$ is degenerate, then it is isotropic and abelian; and*

(3) *$I^\perp$ is a maximal ideal.*

PROOF: Let $I \in \mathfrak{g}$ be any ideal. Then so are $I^\perp$ and $I \cap I^\perp \subset I$, since the intersection of two ideals is an ideal. Since $I$ is minimal, $I \cap I^\perp$ is either $0$ or $I$.

(1) Let's take the first possibility: $I \cap I^\perp = 0$. Definition 2.3 tells us that $I$ is nondegenerate. Since both $I$ and $I^\perp$ are ideals, $[I, I^\perp] \subset I$ and $[I, I^\perp] \subset I^\perp$; hence $[I, I^\perp] \subset I \cap I^\perp = 0$. This means that $[I, I^\perp] = 0$ and $\mathfrak{g} = I \times I^\perp$. Since $I$ is a factor, any ideal of $I$ is automatically an ideal of $\mathfrak{g}$. But by minimality, $I$ cannot have any proper ideals, hence $I$ is either simple or one-dimensional.

(2) The other possibility is that $I \cap I^\perp = I$, which means that $I$ is degenerate. In fact, Definition 2.3 tells us that $I \subset I^\perp$ is isotropic. And by Lemma 2.4, $I \subset I^\perp \subset Z_\mathfrak{g}(I)$, whence it is abelian.

(3) Finally suppose that there exists a proper ideal $J$ such that $I^\perp \subsetneq J$. Taking $\perp$, we find $J^\perp \subsetneq I^{\perp\perp} = I$, which violates minimality. Hence $I^\perp$ is maximal. □

### §3 THE STRUCTURE THEOREM OF MEDINA AND REVOY

The class $\mathcal{C}$ of self-dual Lie algebras is closed under the operation of orthogonal direct product; indeed, if $(\mathfrak{g}_1, \langle-,-\rangle_1)$ and $(\mathfrak{g}_2, \langle-,-\rangle_2)$ are two self-dual Lie algebras, so is $(\mathfrak{g}_1 \times \mathfrak{g}_2, \langle-,-\rangle_1 \oplus \langle-,-\rangle_2)$. We call a self-dual Lie algebra $(\mathfrak{g}, \langle-,-\rangle)$ *indecomposable* if it cannot be written as such a direct product; and *decomposable* if it can. The following preliminary result on the structure of self-dual Lie algebras follows immediately from Proposition 2.6.

COROLLARY 3.1. *Let $(\mathfrak{g}, \langle-,-\rangle)$ be an indecomposable Lie algebra. Then exactly one of the following cases hold:*

(1) *$\mathfrak{g}$ is simple,*

(2) *$\mathfrak{g}$ is one-dimensional, or*

(3) *$\mathfrak{g}$ is not simple, $\dim \mathfrak{g} > 1$, and every proper ideal of $\mathfrak{g}$ is degenerate.* □

It is clear that every self-dual Lie algebra is a product of objects of the types described above. Objects of types (1) and (2) are well-known: they correspond to the direct product of a semisimple Lie algebra and an abelian Lie algebra; that is, they are reductive. The class of objects of type (3) is more exotic. We will see that such indecomposable self-dual Lie algebras are easy to characterise. We shall do so in steps.

Let $(\mathfrak{d}, \langle -, - \rangle) \in \mathcal{C}$ be indecomposable with $\mathfrak{d}$ not simple and with $\dim \mathfrak{d} > 1$. We fix a proper minimal ideal $I \subset \mathfrak{d}$. By Corollary 3.1, it is degenerate and by Proposition 2.6 (2) it is isotropic and abelian. By Lemma 2.4, $I$ is a central ideal of $I^\perp$; whence $\mathfrak{g} = I^\perp/I$ is a Lie algebra. Moreover since $I = I^{\perp\perp}$, $\mathfrak{g}$ inherits an invariant metric $\langle -, - \rangle_\mathfrak{g}$.[1] In other words, we have proven that the following lemma:

LEMMA 3.2. *We have an exact sequence*

$$0 \longrightarrow I \longrightarrow I^\perp \longrightarrow \mathfrak{g} \longrightarrow 0 \qquad (3.3)$$

*with $(\mathfrak{g}, \langle -, - \rangle_\mathfrak{g})$ an object in $\mathcal{C}$.* □

It may seem overkill to use the language of exact sequences, but it turns out that it will be very useful and it will save us some time in the end. For the reader not familiar with this language, we simply recall that a sequence is exact whenever the kernel of any arrow is the image of the following one. In particular, the exactness of the above sequence simply says that $I \subset I^\perp$ is a Lie subalgebra (and an ideal since it is the kernel of a homomorphism) and that $\mathfrak{g} \cong I^\perp/I$ as Lie algebras. Moreover since we are quotienting $I^\perp$ by its $\perp$, the resulting quotient inherits a nondegenerate metric.

Continuing with the argument, by Proposition 2.6 (3), $I^\perp$ is a maximal ideal of $\mathfrak{d}$; whence $\mathfrak{h} = \mathfrak{d}/I^\perp$ is a Lie algebra without proper ideals; that is, either $\mathfrak{h}$ is simple or one-dimensional. It now follows that we can actually identify $\mathfrak{h}$ with a *subalgebra* of $\mathfrak{d}$. The proof of this lemma is much more technical than the proof of the structure theorem and so we leave leave it to the appendix. It is worth pointing out that it is precisely this lemma which prevents the straight-forward extension of the structure theorem to self-dual Lie superalgebras.

LEMMA 3.4. *The exact sequence*

$$0 \longrightarrow I^\perp \longrightarrow \mathfrak{d} \longrightarrow \mathfrak{h} \longrightarrow 0 \;, \qquad (3.5)$$

*splits whenever $\mathfrak{h}$ is simple or one-dimensional; that is, for such $\mathfrak{h}$, $\mathfrak{d} \cong \mathfrak{h} \ltimes I^\perp$.*
□

---

[1] Let us make the following notational remark. Below we will find it necessary to distinguish the Lie bracket and the inner product of the same pair of vectors when thought of as elements of different Lie algebraic structures on the same vector space. We will assume that when nothing is specified, the Lie bracket and the metric correspond to those in $\mathfrak{d}$.

In particular, $\mathfrak{d} = \mathfrak{h} \oplus I^\perp$, whence $V \equiv \mathfrak{h} \oplus I \subset \mathfrak{d}$ is a nondegenerate subspace of $\mathfrak{d}$. This implies that as vector spaces $\mathfrak{d} = V \oplus V^\perp$, and $I^\perp = V^\perp \oplus I$. The map $I^\perp \to \mathfrak{g}$ in (3.3) defines a vector space isomorphism $\mathfrak{g} \cong V^\perp$. With some abuse of notation we will identify $\mathfrak{g}$ with $V^\perp \subset \mathfrak{d}$. In general, though, $V^\perp$ will fail to be a subalgebra. But all that happens is that it acquires a central extension. In fact, if $x, y \in V^\perp$, then

$$[x,y] = [x,y]_\mathfrak{g} + \beta(x,y)$$

for some 2-cocycle $\beta : \bigwedge^2 \mathfrak{g} \to I$, where since $I \subset I^\perp$ is central, it becomes a trivial $\mathfrak{g}$-module. The cohomology class of this cocycle is the one defining the central extension (3.3).

LEMMA 3.6. *$\mathfrak{h}$ acts on $\mathfrak{g}$ via antisymmetric derivations; that is, it preserves the bracket and the invariant metric.*

PROOF: We can define an action of $\mathfrak{h}$ on $\mathfrak{g}$ as follows. Take $x \in \mathfrak{g}$ and lift it to $V^\perp \subset \mathfrak{d}$. If $h \in \mathfrak{h}$, then we can define $h \cdot x = [h,x]$. A priori, this bracket is in $I^\perp = V^\perp \oplus I$. But we show that it is in fact in $V^\perp$. Indeed, let $h' \in \mathfrak{h}$ and compute $\langle h', [h,x]\rangle = \langle [h',h], x\rangle$. Since $[h',h] \in \mathfrak{h} \subset V$, the RHS is zero for all $h'$. Therefore $[h,x] \in \mathfrak{h}^\perp$. But nondegeneracy of $V$ implies that $I \cap \mathfrak{h}^\perp = 0$, whence $[h,x] \in V^\perp$. Since the bracket and metric of $\mathfrak{g}$ are induced from those of $\mathfrak{d}$, and $\mathfrak{h} \in \mathfrak{d}$ acts via (inner) antisymmetric derivations, it also acts on $\mathfrak{g}$ as antisymmetric derivations. □

The action of $\mathfrak{h}$ on $\mathfrak{g}$ is intimately linked to the cocycle $\beta$ characterizing the central extension (3.3). In fact, let $h \in \mathfrak{h}$ and $x, y \in \mathfrak{g}$. Lifting $x$ and $y$ back to $V^\perp \subset \mathfrak{d}$ we compute $\langle h \cdot x, y\rangle_\mathfrak{g} = \langle [h,x], y\rangle = \langle h, [x,y]\rangle = \langle h, [x,y]_\mathfrak{g} + \beta(x,y)\rangle$. But since $\mathfrak{h}$ is orthogonal to $V^\perp$, we find that

$$\langle h \cdot x, y\rangle_\mathfrak{g} = \langle h, \beta(x,y)\rangle \ . \tag{3.7}$$

LEMMA 3.8. *As $\mathfrak{h}$-modules, $I \cong \mathfrak{h}^*$.*

PROOF: Let $h \in \mathfrak{h}$ and $x \in I$. Since $I$ is an ideal, $[h,x] \in I$, whence it is an $\mathfrak{h}$-module. Because $V = \mathfrak{h} \oplus I$ is nondegenerate and $I$ is isotropic, we can identify $I$ with $\mathfrak{h}^*$ as vector spaces: the isomorphism given by $I \ni x \mapsto \langle x, -\rangle$ restricted to $\mathfrak{h}$. Now notice that for all $h' \in \mathfrak{h}$, $\langle [h,x], h'\rangle = -\langle x, [h,h']\rangle$, whence $I \cong \mathfrak{h}^*$ as $\mathfrak{h}$-modules. □

In summary, we have proven the following structure theorem.

THEOREM 3.9. *(Medina–Revoy [1]) Let $(\mathfrak{d}, \langle -, -\rangle)$ be an indecomposable object in $\mathcal{C}$ such that $\mathfrak{d}$ is not simple nor one-dimensional. Then $\mathfrak{d}$ is isomorphic to the Lie algebra with underlying vector space $\mathfrak{g} \oplus \mathfrak{h} \oplus \mathfrak{h}^*$, where*

(1) $\mathfrak{g}$ *is the Lie algebra defined by* (3.3) *and which inherits an invariant metric* $\langle -,-\rangle_\mathfrak{g}$ *by restricting* $\langle -,-\rangle$;

(2) $\mathfrak{h}$ *is the Lie algebra defined by* (3.5), *which is either one-dimensional or simple*;

(3) $\mathfrak{h}$ *acts on* $\mathfrak{g}$ *via antisymmetric derivations*: $(h,x) \mapsto h \cdot x$ *for* $h \in \mathfrak{h}$ *and* $x \in \mathfrak{g}$; *and*

(4) *the Lie brackets are given for* $x, x' \in \mathfrak{g}$, $h, h' \in \mathfrak{h}$ *and* $\alpha, \alpha' \in \mathfrak{h}^*$ *by*

$$[(x,h,\alpha),(x',h',\alpha')] = \\ ([x,x']_\mathfrak{g} + h\cdot x' - h'\cdot x, [h,h']_\mathfrak{h}, \beta(x,x') + \mathrm{ad}^*_h\cdot\alpha' - \mathrm{ad}^*_{h'}\cdot\alpha) , \tag{3.10}$$

*where* $\beta : \bigwedge^2 \mathfrak{g} \to \mathfrak{h}^*$ *is given by* (3.7). $\square$

## §4 Double Extensions

In this section we review the definition of a double extension, formalising the results in the previous section on the structure of nonreductive indecomposable self-dual Lie algebras.

DEFINITION 4.1. Given $(\mathfrak{g}, \langle -,-\rangle_\mathfrak{g}) \in \mathcal{C}$ and $\mathfrak{h}$ a Lie algebra acting on $\mathfrak{g}$ via antisymmetric derivations, the Lie algebra $\mathfrak{d}$ defined on the vector space $\mathfrak{g} \oplus \mathfrak{h} \oplus \mathfrak{h}^*$ by (3.10) and (3.7) is called the *double extension of* $\mathfrak{g}$ *by* $\mathfrak{h}$ and we denote it by $D(\mathfrak{g},\mathfrak{h})$.

REMARK 4.2. Notice that the notation $D(\mathfrak{g},\mathfrak{h})$ is ambiguous on two counts. First of all, the data is not just $\mathfrak{g}$ but $(\mathfrak{g}, \langle -,-\rangle_\mathfrak{g})$; and also not just $\mathfrak{h}$ but $\mathfrak{h}$ together with the action of $\mathfrak{h}$ on $\mathfrak{g}$ via antisymmetric derivations. Nevertheless, for the purposes of this paper, any ambiguity that might arise will be resolved contextually.

The double extension of $\mathfrak{g}$ by $\mathfrak{h}$ always carries an invariant metric. Indeed, if $\langle -,-\rangle_\mathfrak{h}$ is any invariant bilinear form on $\mathfrak{h}$, we define

$$\langle (x,h,\alpha),(x',h',\alpha')\rangle = \langle x, x'\rangle_\mathfrak{g} + \langle h, h'\rangle_\mathfrak{h} + \alpha(h') + \alpha'(h) , \tag{4.3}$$

for all $x, x' \in \mathfrak{g}$, $h, h' \in \mathfrak{h}$ and $\alpha, \alpha' \in \mathfrak{h}^*$. A routine calculation shows that this metric is invariant, whence $(\mathfrak{d}, \langle -,-\rangle)$ is a self-dual Lie algebra.

REMARK 4.4. It is worth pointing out that if the action of $\mathfrak{h}$ on $\mathfrak{g}$ is trivial, then the double extension is decomposable and isomorphic to $\mathfrak{g} \times (\mathfrak{h} \ltimes \mathfrak{h}^*)$. In particular, if $\mathfrak{g} = 0$, then the double extension is $(\mathfrak{h} \ltimes \mathfrak{h}^*)$, which is the classical double of $\mathfrak{h}$ given the trivial bialgebra structure [6].

It will be convenient later on to work with the explicit expression for the Lie brackets and for the invariant metric of a double extension. Hence we will now work out these expressions in a basis. We let $(\mathfrak{g}, \langle -, - \rangle_\mathfrak{g})$ be a self-dual Lie algebra. Let the invariant metric have components $\Omega^\mathfrak{g}_{ij}$ relative to a fixed basis $\{G_i\}$. Also relative to this basis, we will let the Lie bracket in $\mathfrak{g}$ be given by $[G_i, G_j]_\mathfrak{g} = f_{ij}{}^k G_k$. We consider also a Lie algebra $\mathfrak{h}$, with basis $\{H_\alpha\}$, acting on $\mathfrak{g}$ via antisymmetric derivations

$$H_\alpha \cdot G_i = f_{\alpha i}{}^j G_j ,$$

and with Lie brackets given by $[H_\alpha, H_\beta]_\mathfrak{h} = f_{\alpha\beta}{}^\gamma H_\gamma$. Its dual $\mathfrak{h}^*$ has canonical dual basis given by $\{H^\alpha\}$.

The double extension $\mathfrak{d} = D(\mathfrak{g}, \mathfrak{h})$ will be then defined on the vector space $\mathfrak{g} \oplus \mathfrak{h} \oplus \mathfrak{h}^*$ by the following Lie brackets

$$[G_i, G_j] = f_{ij}{}^k G_k + f_{ij\alpha} H^\alpha ,$$
$$[H_\alpha, G_i] = f_{\alpha i}{}^j G_j ,$$
$$[H_\alpha, H_\beta] = f_{\alpha\beta}{}^\gamma H_\gamma ,$$
$$[H_\alpha, H^\beta] = -f_{\alpha\gamma}{}^\beta H^\gamma ,$$
$$[H^\alpha, G_i] = [H^\alpha, H^\beta] = 0 ,$$

where $f_{ij\alpha} = f_{\alpha i}{}^k \Omega^\mathfrak{g}_{kj}$. The above explicit expression, makes manifest the fact mentioned in Remark 4.2 that $D(\mathfrak{g}, \mathfrak{h})$ does not depend on $\mathfrak{g}$ and $\mathfrak{h}$ only through their Lie algebra structures, as the notation would suggest, but also on the action of $\mathfrak{h}$ on $\mathfrak{g}$ and on the metric of $\mathfrak{g}$.

The invariant metric on $D(\mathfrak{g}, \mathfrak{h})$ is given by

$$\Omega^\mathfrak{d}_{ab} = \begin{array}{c} \\ G_i \\ H_\alpha \\ H^\alpha \end{array} \begin{pmatrix} G_j & H_\beta & H^\beta \\ \Omega^\mathfrak{g}_{ij} & 0 & 0 \\ 0 & h_{\alpha\beta} & \delta_\alpha{}^\beta \\ 0 & \delta^\alpha{}_\beta & 0 \end{pmatrix} ,$$

where $(h_{\alpha\beta})$ is an arbitrary (possibly degenerate) invariant bilinear form in $\mathfrak{h}$. We also record for future use the Killing form of the above double extension $\kappa^\mathfrak{d}$. This form will of course be degenerate, having the form

$$\kappa^\mathfrak{d}_{ab} = \begin{array}{c} \\ G_i \\ H_\alpha \\ H^\alpha \end{array} \begin{pmatrix} G_j & H_\beta & H^\beta \\ \kappa^\mathfrak{g}_{ij} & \kappa^\mathfrak{d}_{i\beta} & 0 \\ \kappa^\mathfrak{d}_{\alpha j} & \kappa^\mathfrak{d}_{\alpha\beta} & 0 \\ 0 & 0 & 0 \end{pmatrix} ,$$

where $\kappa^\mathfrak{g}$ is the Killing form of $\mathfrak{g}$ and where

$$\kappa^\mathfrak{d}_{i\alpha} = f_{ij}{}^k f_{\alpha k}{}^j \quad \text{and} \quad \kappa^\mathfrak{d}_{\alpha\beta} = f_{\alpha i}{}^j f_{\beta j}{}^i + 2\kappa^\mathfrak{h}_{\alpha\beta} .$$

## §5 Some useful refinements

In this section we prove some refinements of Theorem 3.9 which have proven instrumental in the applications to Conformal Field Theory. We start by listing some conditions on $\mathfrak{g}$ under which any double extension $D(\mathfrak{g}, \mathfrak{h})$ will fail to be indecomposable.

Remark 4.4 tells us that a double extension need not be indecomposable even if $\mathfrak{h}$ is taken to be simple or one-dimensional; and one such example is when $\mathfrak{h}$ acts on $\mathfrak{g}$ via inner derivations, as we now see.

PROPOSITION 5.1. *If $\mathfrak{h}$ acts on $(\mathfrak{g}, \langle -, - \rangle) \in \mathcal{C}$ via inner derivations, then $D(\mathfrak{g}, \mathfrak{h}) \cong \mathfrak{g} \times (\mathfrak{h} \ltimes \mathfrak{h}^*)$.*

PROOF: Let $\varphi : \mathfrak{h} \to \mathfrak{g}$ be the homomorphism defining the action of $\mathfrak{h}$ on $\mathfrak{g}$. In other words, for $h \in \mathfrak{h}$ and $x \in \mathfrak{g}$, $h \cdot x = [\varphi(h), x]_{\mathfrak{g}}$. Let $\varphi^\flat : \mathfrak{h} \to \mathfrak{g}^*$ be the map sending $h \mapsto \langle \varphi(h), - \rangle_{\mathfrak{g}}$, and let $\varphi^\sharp : \mathfrak{g} \to \mathfrak{h}^*$ denote its transpose. Notice that because the metric of $\mathfrak{g}$ is $\mathfrak{h}$-invariant, these maps are actually intertwiners of the action of $\mathfrak{h}$; in particular, for $h \in \mathfrak{h}$ and $x \in \mathfrak{g}$, we have that

$$[h, \varphi^\sharp(x)] = \varphi^\sharp([\varphi(h), x]_{\mathfrak{g}}) \ . \tag{5.2}$$

We can now define the following vector space automorphism $\Psi$ of $\mathfrak{g} \oplus \mathfrak{h} \oplus \mathfrak{h}^*$:

$$\Psi(x, h, \alpha) = (x - \varphi^\sharp(x), h + \varphi(h), \alpha) \ .$$

We claim that $\Psi$ is a Lie algebra isomorphism $D(\mathfrak{g}, \mathfrak{h}) \xrightarrow{\cong} \mathfrak{g} \times (\mathfrak{h} \ltimes \mathfrak{h}^*)$. Indeed, for $x, y \in \mathfrak{g}$, $\Psi([x,y]) = \Psi([x,y]_{\mathfrak{g}} + \beta(x,y)) = [x,y]_{\mathfrak{g}} - \varphi^\sharp([x,y]_{\mathfrak{g}}) + \beta(x,y)$. But from (3.7) we have that $\beta(x,y) = \varphi^\sharp([x,y]_{\mathfrak{g}})$, whence $\Psi([x,y]) = [\Psi(x), \Psi(y)]$. Secondly, we have that on the one hand, for $h \in \mathfrak{h}$, $\Psi([h,x]) = \Psi([\varphi(h),x]_{\mathfrak{g}}) = [\varphi(h),x]_{\mathfrak{g}} - \varphi^\sharp([\varphi(h),x]_{\mathfrak{g}})$ and $[\Psi(h), \Psi(x)] = [h + \varphi(h), x - \varphi^\sharp(x)] = [\varphi(h), x]_{\mathfrak{g}} - [h, \varphi^\sharp(x)]$ on the other. But both of these expressions agree by virtue of (5.2). Similarly, $\Psi([h,h']) = [h,h'] + \varphi([h,h'])$ agrees with $[\Psi(h), \Psi(h')] = [h+\varphi(h), h'+\varphi(h')] = [h,h'] + [\varphi(h), \varphi(h')]$ for all $h' \in \mathfrak{h}$, since $\varphi$ is a homomorphism of Lie algebras. The rest of the brackets are verified in a similar fashion. □

REMARK 5.3. The invariant metric in $(\mathfrak{h} \ltimes \mathfrak{h}^*)$ is now given by

$$\langle (h, \alpha), (h', \alpha') \rangle = \langle h, h' \rangle_{\mathfrak{h}} + \langle \varphi(h), \varphi(h') \rangle_{\mathfrak{g}} + \alpha(h') + \alpha'(h) \ .$$

In other words, $\langle -, - \rangle_{\mathfrak{h}}$ receives a correction coming from the pull-back by $\varphi$ to $\mathfrak{h}$ of the invariant metric in $\mathfrak{g}$; that is, $\langle -, - \rangle_{\mathfrak{h}} + \varphi^* \langle -, - \rangle_{\mathfrak{p}}$.

In particular, if all the antisymmetric derivations of $\mathfrak{g}$ are inner, then $\mathfrak{g}$ factors out of the double extension. This idea can be pursued further, but first a definition. Recall a (real) Lie algebra $\mathfrak{g}$ is *perfect* if $[\mathfrak{g}, \mathfrak{g}] = \mathfrak{g}$ or, equivalently, if $H^1(\mathfrak{g}; \mathbb{R}) = 0$. By analogy let us define the following.

DEFINITION 5.4. We say that a (real) Lie algebra $\mathfrak{p}$ is *pluperfect* whenever $H^1(\mathfrak{p}; \mathbb{R}) = H^2(\mathfrak{p}; \mathbb{R}) = 0$. Notice that semisimple Lie algebras are pluperfect.

THEOREM 5.5. *The Lie algebra $\mathfrak{g}$ in Theorem 3.9 cannot have a pluperfect factor.*

PROOF: We will prove that if $\mathfrak{g}$ has a pluperfect factor, then its double extension is decomposable, in contradiction to the hypothesis of Theorem 3.9. Thus let $(\mathfrak{g}, \langle -, - \rangle_{\mathfrak{g}})$ be an object in $\mathcal{C}$ such that $\mathfrak{g} = \mathfrak{p} \times \mathfrak{a}$ with $\mathfrak{p}$ pluperfect and $\mathfrak{a}$ arbitrary without pluperfect factors.

(1) $\mathfrak{p}$ *and* $\mathfrak{a}$ *are orthogonal.*

$\mathfrak{p}$ is in particular perfect, which together with the invariance of the metric implies that $\langle \mathfrak{p}, \mathfrak{a} \rangle = \langle [\mathfrak{p}, \mathfrak{p}], \mathfrak{a} \rangle = \langle \mathfrak{p}, [\mathfrak{p}, \mathfrak{a}] \rangle = 0$.

(2) *Let* $\mathrm{Der}_a$ *stand for the antisymmetric derivations. Then* $\mathrm{Der}_a \mathfrak{g} = \mathfrak{p} \times \mathrm{Der}_a \mathfrak{a}$.

Let $d \in \mathrm{Der}_a \mathfrak{g}$ be an antisymmetric derivation. If $x \in \mathfrak{g}$ we write $d(x) = d_1(x) + d_2(x)$ where $d_1(x) \in \mathfrak{p}$ and $d_2(x) \in \mathfrak{a}$. Let $s, s' \in \mathfrak{p}$. Since $d$ is a derivation, we have $[d(s), s'] + [s, d(s')] = d([s, s'])$. Breaking it up into its components, we find that $[d_1(s), s'] + [s, d_1(s')] = d_1([s, s'])$ and that $d_2([s, s']) = 0$. The former equation says that $d_1 \in \mathrm{Der}\,\mathfrak{p}$, whereas the latter says that $d_2$ annihilates $[\mathfrak{p}, \mathfrak{p}] = \mathfrak{p}$. If $a \in \mathfrak{a}$ we have $[s, d(a)] + [d(s), a] = 0$. Breaking it up into components we find $[s, d_1(a)] = 0$, which says that $d_1(a)$ is central in $\mathfrak{p}$. But since $[\mathfrak{p}, \mathfrak{p}] = \mathfrak{p}$, Lemma 2.5 says that the center is trivial, whence $d_1(a)$ must vanish. If $a' \in \mathfrak{a}$, then $[d(a), a'] + [a, d(a')] = d([a, a'])$, which breaks up as $[d_2(a), a'] + [a, d_2(a')] = d_2([a, a'])$. This means that $d_2 \in \mathrm{Der}\,\mathfrak{a}$. Finally, the antisymmetry condition says that $d_2 \in \mathrm{Der}_a \mathfrak{a}$ whereas from $H^2(\mathfrak{p}; \mathbb{R}) = 0$ it follows that all antisymmetric derivations of $\mathfrak{p}$ are inner: every antisymmetric derivation $D \in \mathrm{Der}_a \mathfrak{p}$ defines a 2-cocycle by $\gamma(s, s') = \langle d(s), s' \rangle$ which is a coboundary $\gamma(s, s') = -\eta([s, s'])$, for some $\eta \in \mathfrak{p}^*$. But this means that there exists $s'' \in \mathfrak{p}$ such that $-\eta([s, s']) = \langle s'', [s, s'] \rangle = \langle [s'', s], s' \rangle$, whence $d(s) = [s'', s]$ is inner. Conversely, all inner derivations are antisymmetric, so that $d_1 \in \mathrm{ad}\,\mathfrak{p}$.

In particular, since $\mathfrak{h}$ acts on $\mathfrak{g}$ via inner derivations, there exists a Lie algebra (hence $\mathfrak{h}$-module) morphism $\varphi : \mathfrak{h} \to \mathfrak{p}$ such that for $h \in \mathfrak{h}$ and $s \in \mathfrak{p}$, $h \cdot s = [\varphi(h), s]_{\mathfrak{p}}$. Then the proof of Proposition 5.1 implies, *mutatis mutandis*, that $\mathfrak{p}$ factors out of the double extension.

In other words, the vector space automorphism $\Psi$ of $\mathfrak{p} \oplus \mathfrak{a} \oplus \mathfrak{h} \oplus \mathfrak{h}^*$:

$$\Psi(s, a, h, \alpha) = (s - \varphi^\sharp(s), a, h + \varphi(h), \alpha) .$$

defines a Lie algebra isomorphism $D(\mathfrak{p} \times \mathfrak{a}, \mathfrak{h}) \xrightarrow{\simeq} \mathfrak{p} \times D(\mathfrak{a}, \mathfrak{h})$. Furthermore, the invariant metric in $D(\mathfrak{a}, \mathfrak{h})$ is now given by

$$\begin{aligned}
\langle (a, h, \alpha) , (a', h', \alpha') \rangle = \\
\langle a , a' \rangle_\mathfrak{a} + \langle h , h' \rangle_\mathfrak{h} + \langle \varphi(h) , \varphi(h') \rangle_\mathfrak{p} + \alpha(h') + \alpha'(h) .
\end{aligned}$$

This concludes the proof of Theorem 5.5. □

As a corollary of Theorem 3.9 and Theorem 5.5 we have the following characterisation of the class $\mathcal{C}$.

COROLLARY 5.6. *The class $\mathcal{C}$ breaks up as $\mathcal{C}_S \times \mathcal{C}_N$ where $\mathcal{C}_S$ is the subclass of semisimple Lie algebras and $\mathcal{C}_N$ is the smallest class of real finite-dimensional Lie algebras containing the one-dimensional Lie algebra and closed under the operations of direct product and double extension by a simple or one-dimensional algebra. In particular, all objects in $\mathcal{C}_N$ are nonsemisimple. Moreover the subclass $\mathcal{C}'_S \subset \mathcal{C}_N$ of solvable Lie algebras is class generated by the one-dimensional Lie algebra under the operations of direct product and double extension by the one-dimensional Lie algebra.* □

## §6 Deforming the invariant metric

Let $(\mathfrak{d}, \Omega)$ be a self-dual Lie algebra and let $\kappa$ denote its Killing form. We would to ask whether one can deform the metric $\Omega$ while retaining nondegeneracy. Rather than analyse this problem in full generality, we will limit ourselves to the case of interest in Conformal Field Theory. Namely, we will deform $\Omega$ by a scalar multiple of the Killing form. Such shifts are the typical effect of quantum renormalisation. Let $t$ be a scalar (a real or complex number) and let $g_t$ denote the bilinear form $g_t = \Omega - t\kappa$. Fix $t$ once and for all and define $\mathfrak{d}^\perp$ to be the radical of $g_t$; that is $\mathfrak{d}^\perp = \{v \in \mathfrak{d} | g_t(v, w) = 0 \ \forall w \in \mathfrak{g}\}$. Notice that $\mathfrak{d}^\perp \subset \mathfrak{d}$ is an ideal, since the bilinear form $g_t$ is invariant. In particular, $\mathfrak{d}^\perp$ is a Lie algebra. We will prove the following result:

THEOREM 6.1. *If $(\mathfrak{d}, \Omega)$ is an indecomposable self-dual Lie algebra, then $\mathfrak{d}^\perp = 0$ unless $\mathfrak{d}$ is simple and $\Omega = t\kappa$, in which case $g_t = 0$ and $\mathfrak{d}^\perp = \mathfrak{d}$.*

PROOF: Since $(\mathfrak{d}, \Omega)$ is indecomposable, then by Theorem 3.9 it is either simple, one-dimensional, or a double extension $D(\mathfrak{g}, \mathfrak{h})$ where $\mathfrak{h}$ is simple or one-dimensional. The theorem is clear for the first two cases, as we now show. If $\mathfrak{d}$ is one-dimensional, then $\kappa = 0$ and $\mathfrak{d}^\perp = 0$; and similarly if $\mathfrak{d}$ is simple, then $\mathfrak{d}^\perp$, being an ideal, must be either 0 or $\mathfrak{d}$; the latter case corresponding to the case $g_t = 0$, or equivalently $\Omega = t\kappa$. Therefore all we have left to tackle is the case where $\mathfrak{d} = D(\mathfrak{g}, \mathfrak{h})$ is a double extension. The theorem will follow if we can prove that $\mathfrak{d}^\perp = 0$ in this case.

We proceed by induction on the dimension of the Lie algebra. Suppose that the theorem is true for all indecomposable self-dual Lie algebras of dimension $\leq N$—the case $N = 1$ being trivially satisfied—and let $\mathfrak{d} = D(\mathfrak{g}, \mathfrak{h})$ be an indecomposable double-extension of dimension[2] $N + 1$. The theorem follows if we can prove that $\mathfrak{d}^\perp = 0$. We now have the following lemma, whose proof we give below:

LEMMA 6.2. *Let $\mathfrak{d} = D(\mathfrak{g}, \mathfrak{h})$ be a double extension, with $\mathfrak{g}$ and $\mathfrak{h}$ arbitrary. Then there is a Lie algebra isomorphism:*

$$\mathfrak{d}^\perp = D(\mathfrak{g}, \mathfrak{h})^\perp \cong \mathfrak{g}^\perp .$$

□

Using Lemma 6.2, we have that $\mathfrak{d}^\perp \cong \mathfrak{g}^\perp$. In general $\mathfrak{g}$ need not be indecomposable, so write it as $\mathfrak{g} = \mathfrak{g}_1 \times \cdots \times \mathfrak{g}_k$, where each $\mathfrak{g}_i$ is indecomposable. Clearly, $\mathfrak{g}^\perp \cong \mathfrak{g}_1^\perp \times \cdots \times \mathfrak{g}_k^\perp$. Since $\dim \mathfrak{g}_i < \dim \mathfrak{d}$ for each $i$, we can apply the induction hypothesis to deduce that $\mathfrak{g}_i^\perp$ will only be nonzero when $\mathfrak{g}_i$ is simple. But if $\mathfrak{g}$ would have a simple factor, Theorem 5.5 would imply that $\mathfrak{d}$ is decomposable, violating the hypothesis. Therefore $\mathfrak{g}^\perp = 0$ and we can extend the induction hypothesis. □

COROLLARY 6.3. *Let $(\mathfrak{d}, \Omega)$ be any self-dual Lie algebra. Then*

*(1) $\mathfrak{d}^\perp$ is semisimple;*

*(2) $\mathfrak{d}$ decomposes into an orthogonal direct sum $\mathfrak{d} = \mathfrak{d}^\perp \times \mathfrak{d}_1$, where $\mathfrak{d}^\perp$ is semisimple, and $\mathfrak{d}_1^\perp = 0$.*

PROOF: (1) is an immediate corollary. Since $\mathfrak{d}^\perp$ is a semisimple ideal, it is a factor; hence (2). □

Finally we prove the lemma.

PROOF: (of Lemma 6.2) In the explicit basis introduced in Section 4, we let $v = v^j G_j + v^\alpha H_\alpha + v_\alpha H^\alpha$ belong to $\mathfrak{d}^\perp$ and let us see what this implies. The bilinear form defining $\perp$ is $\Omega_{ab}^{\mathfrak{d}} - t\kappa_{ab}^{\mathfrak{d}}$, whose matrix is given by

$$\begin{array}{c} \phantom{G_i} \\ G_i \\ H_\alpha \\ H^\alpha \end{array} \begin{array}{c} G_j \phantom{XXX} H_\beta \phantom{XXX} H^\beta \\ \begin{pmatrix} \Omega_{ij}^{\mathfrak{g}} - t\kappa_{ij}^{\mathfrak{g}} & -t\kappa_{i\beta}^{\mathfrak{d}} & 0 \\ -t\kappa_{\alpha j}^{\mathfrak{d}} & h_{\alpha\beta} - t\kappa_{\alpha\beta}^{\mathfrak{d}} & \delta_\alpha{}^\beta \\ 0 & \delta^\alpha{}_\beta & 0 \end{pmatrix} \end{array}.$$

---

[2] Actually we have not shown that there is an indecomposable self-dual Lie algebra in every dimension. So if there is no indecomposable $\mathfrak{d} = D(\mathfrak{g}, \mathfrak{h})$ in dimension $N+1$ then take one of the smallest dimension $> N$.

Therefore, $v \in \mathfrak{d}^\perp$ implies that

$$(\Omega_{ab}^{\mathfrak{d}} - t\kappa_{ab}^{\mathfrak{d}}) \begin{pmatrix} v^j \\ v^\beta \\ v_\beta \end{pmatrix} = \begin{pmatrix} (\Omega_{ij}^{\mathfrak{g}} - t\kappa_{ij}^{\mathfrak{g}})v^j - t\kappa_{i\beta}^{\mathfrak{d}}v^\beta \\ -t\kappa_{\alpha j}^{\mathfrak{d}} v^j + (h_{\alpha\beta} - t\kappa_{\alpha\beta}^{\mathfrak{d}})v^\beta + v_\alpha \\ v^\alpha \end{pmatrix} = 0 .$$

This in turn yields the equations $v^\alpha = 0$, $v_\alpha = t\kappa_{\alpha j}^{\mathfrak{d}} v^j$, and

$$(\Omega_{ij}^{\mathfrak{g}} - t\kappa_{ij}^{\mathfrak{g}})v^j = 0 ,$$

whence $v^j G_j$ belongs to $\mathfrak{g}^\perp$. Conversely, any $v^j G_j \in \mathfrak{g}^\perp$ extends to a vector $v^j G_j + t v^j \kappa_{\alpha j}^{\mathfrak{d}} H^\alpha$ which by the above computation belongs to $\mathfrak{d}^\perp$. In summary, we have a vector space isomorphism $s : \mathfrak{g}^\perp \to \mathfrak{d}^\perp$, defined by $s(v^j G_j) = v^j G'_j$, where $G'_j = G_j + t\kappa_{\alpha j}^{\mathfrak{d}} H^\alpha$. We will now show that this is also an isomorphism of Lie algebras. Computing the brackets in $\mathfrak{d}$, we obtain

$$\begin{aligned}{} [s(G_i), s(G_j)]_{\mathfrak{d}} = [G'_i, G'_j]_{\mathfrak{d}} &= f_{ij}{}^k G_k + f_{ij\alpha} H^\alpha \\ &= f_{ij}{}^k G'_k + (f_{ij\alpha} - t f_{ij}{}^k \kappa_{k\alpha}^{\mathfrak{d}}) H^\alpha . \end{aligned} \qquad (6.4)$$

Now notice that $f_{ij\alpha} = f_{ij}{}^a \Omega_{a\alpha}^{\mathfrak{d}}$, and that $f_{ij}{}^k \kappa_{k\alpha}^{\mathfrak{d}} = f_{ij}{}^a \kappa_{a\alpha}^{\mathfrak{d}}$; so that we can rewrite (6.4) as

$$[G'_i, G'_j]_{\mathfrak{d}} = f_{ij}{}^k G'_k + f_{ij}{}^a (g_t^{\mathfrak{d}})_{a\alpha} H^\alpha .$$

Using that $g_t^{\mathfrak{d}}$ is an invariant bilinear form, we arrive at

$$[G'_i, G'_j]_{\mathfrak{d}} = f_{ij}{}^k G'_k - f_{i\alpha}{}^a (g_t^{\mathfrak{d}})_{aj} H^\alpha .$$

Finally we notice that $f_{i\alpha}{}^a (g_t^{\mathfrak{d}})_{aj} = f_{i\alpha}{}^k (g_t^{\mathfrak{d}})_{kj}$ and that the restriction of $g_t^{\mathfrak{d}}$ to $\mathfrak{g}$ coincides with $g_t^{\mathfrak{g}}$, so that we end up with

$$[G'_i, G'_j]_{\mathfrak{d}} = f_{ij}{}^k G'_k - f_{i\alpha}{}^k (g_t^{\mathfrak{d}})_{jk} H^\alpha ,$$

which shows explicitly that if $v^i G_i$ and $w^j G_j$ are in $\mathfrak{g}^\perp$, then

$$[s(v^i G_i), s(w^j G_j)]_{\mathfrak{d}} = s([v^i G_i, w^j G_j]_{\mathfrak{g}}) ,$$

so that $s$ is a homomorphism. $\qquad \square$

## §7  Applications in Conformal Field Theory

To conclude we would like to mention very briefly some of the applications of self-dual Lie algebras and, in particular, of the results in this paper in Conformal Field Theory. We will be brief and will limit ourselves mostly to directing the attention of the reader to the relevant literature.

Any conformally invariant two-dimensional $\sigma$-model (of the right central charge) is a possible (bosonic) string background. Any such $\sigma$-model is classically conformally invariant, but demanding that this persists upon quantisation imposes equations on the metric which need to be satisfied. Therefore in the process of satisfying these equations, the classical form of the metric gets "renormalised" and it is seldom the case that one can write down a nontrivial metric which exactly solves the equations–that is, an exact string background. In [7] Nappi and Witten constructed one such exact string background out of a WZW model with target space a four-dimensional solvable Lie group. The nonperturbative proof of the conformal invariance of the theory made use of a Sugawara construction built out of a nondegenerate metric on the Lie algebra. This fact prompted Mohammedi [2] to investigate the existence of a Sugawara construction on a given Lie algebra $\mathfrak{g}$. The conclusion of his analysis (see also [3]) is that a Sugawara construction exists if and only if $\mathfrak{g}$ is self-dual. Self-dual Sugawara constructions have appeared also in the work of Lian [5] on finitely-generated simple vertex operator algebras.

In [3] we analysed the Sugawara constructions arising out of self-dual Lie algebras, with the motivation of answering the following question. All examples of nonsemisimple Sugawara constructions known at the time shared the property that the central charge was equal to the dimension of the Lie algebra. *Was this inevitable or could one construct CFTs with non-integral values of the central charge?* The answer turns out to be negative—a fact we established in [3]. More precisely, we used Corollary 5.6 (derived from a weaker version of Theorem 5.5) to deduce that the Sugawara central charge associated with any self-dual Lie algebra in the class $\mathcal{C}_N$ is integral and equal to the dimension of the Lie algebra. Hence any other value for the central charge has its origins in a Sugawara construction in the class $\mathcal{C}_S$. A detailed analysis of the self-dual Lie algebras in low dimension has been made by Kehagias, who classified the WZW models in (target) dimension $\leq 5$ and, in addition, all those six-dimensional ones with nilpotent target group [8]. More recently, we have studied the coset constructions arising out of self-dual Lie algebras [4] as well as settled some issues concerning the relation between the WZW model and the Sugawara construction, for which Theorem 6.1 proved instrumental. In [4] the reader may also find more references on nonreductive Sugawara constructions and (gauged) WZW models.

We conclude this brief survey of applications with a word on supersymmetry. It was proven in [2] that the condition for the existence of an $N=1$ supersymmetric Sugawara construction on the $N=1$ affine Lie algebra $\widehat{\mathfrak{g}}_{N=1}$ is also that $\mathfrak{g}$ be self-dual. This opens the possibility of studying for which self-dual Lie algebras $\mathfrak{g}$, does the $N=1$ Sugawara construction $\widehat{\mathfrak{g}}_{N=1}$ admit an $N=2$ extension. The conditions were found in [9] and re-interpreted in [10] in order to classify all those solvable six-dimensional self-dual Lie algebras admitting an $N=2$ construction. The condition on the dimensionality is motivated by String Theory, since the central charge for a six-dimensional solvable Lie algebra is equal to 9.

§8  SOME OPEN PROBLEMS

We have seen that the results of Medina and Revoy, suitably refined and augmented, have a wide applicability in Conformal Field Theory. Certainly there still remains a lot to be learned from self-dual Lie algebras and, if we compare them with semisimple Lie algebras, very little is known about them indeed. At this stage a complete classification is hard to envision, but some more modest results would be welcome; for instance, the classification of all six-dimensional self-dual Lie algebras. An interesting open problem is the extension of these results to self-dual Lie superalgebras. As shown in [11] and in [4], self-dual Lie superalgebras also lead to Sugawara constructions. Motivated by this fact, one would like to have a structure theorem for such Lie superalgebras. If we study closely the results outlined in this paper, one sees that one can substitute Lie algebra for Lie superalgebra in many of the results and the statements and proofs still hold *mutatis mutandis*. (Statements like Corollary 5.6 would, of course, have to be modified, since not all simple Lie superalgebras are self-dual.) The only exception is Lemma 3.4, for which we have not been able to find a proof nor a counterexample. The notion of a double extension still works and allows one to construct self-dual Lie superalgebras, but without a superanalogue of , nothing guarantees that this is the way to obtain them all.

Notice that although the results described in this paper hold for symmetric self-dual Lie algebras, one could have used an antisymmetric form for much of the discussion in this paper and many of the results would have remained unchanged. However, had we dropped any symmetry requirements whatsoever, the results need severe modification. The determination of a structure theorem for these more general self-dual Lie algebras is an open problem—one we find intriguing and to which we may return to it elsewhere.

Appendix A    Split extensions of some Lie algebras

We now prove Lemma 3.4. More precisely we prove the following result, which implies Lemma 3.4.

PROPOSITION A.1. *If $\mathfrak{c}$ is simple or one-dimensional, every Lie algebra extension*

$$0 \longrightarrow \mathfrak{a} \longrightarrow \mathfrak{b} \longrightarrow \mathfrak{c} \longrightarrow 0 \tag{A.2}$$

*splits.*

REMARK A.3. Before we proceed to the proof we mention that if $\mathfrak{a}$ were abelian the result would follow immediately as a consequence of the second Whitehead lemma: $H^2(\mathfrak{c};\mathfrak{a}) = 0$. Similarly, if $\mathfrak{a}$ were solvable one could proceed by induction on the derived length of $\mathfrak{a}$. If on the other hand, $\mathfrak{a}$ were semisimple, then $\mathfrak{b}$ would decompose as a direct product $\mathfrak{a} \times \mathfrak{c}$. However we are interested in general $\mathfrak{a}$. This result should be standard but we have not found it in the literature.

PROOF: (of Proposition A.1) Let $0 \to \mathfrak{r} \to \mathfrak{a} \to \mathfrak{l} \to 0$ be a Levi decomposition for $\mathfrak{a}$ with $\mathfrak{r}$ the radical and $\mathfrak{l}$ semisimple. Since $\mathfrak{r}$ is a characteristic ideal of $\mathfrak{a}$ and $\mathfrak{a}$ is an ideal of $\mathfrak{b}$, $\mathfrak{r}$ is an ideal of $\mathfrak{b}$. Let $\mathfrak{g} \equiv \mathfrak{b}/\mathfrak{r}$. Since $\mathfrak{l} \subset \mathfrak{a}$ is a subalgebra we have a map $\mathfrak{l} \to \mathfrak{g}$ induced by the composition $\mathfrak{a} \to \mathfrak{b} \to \mathfrak{g}$. It is clear that this map is one-to-one and the image of $\mathfrak{l}$ in $\mathfrak{g}$ is an ideal. Define then $\mathfrak{h} \equiv \mathfrak{g}/\mathfrak{l}$. Since $\mathfrak{l}$ is semisimple, the sequence $0 \to \mathfrak{l} \to \mathfrak{g} \to \mathfrak{h} \to 0$ splits, and we can identify $\mathfrak{h}$ with an ideal of $\mathfrak{g}$, so that $\mathfrak{g} = \mathfrak{h} \times \mathfrak{l}$. Notice that $\dim \mathfrak{c} = \dim \mathfrak{h}$. More is true however.

CLAIM: *As Lie algebras, $\mathfrak{c} \cong \mathfrak{h}$, whence $\mathfrak{g} = \mathfrak{l} \times \mathfrak{c}$.*

To prove this take any (vector space) section $\sigma : \mathfrak{c} \to \mathfrak{b}$ through $\mathfrak{b} \to \mathfrak{c} \to 0$. We define a map $\varphi : \mathfrak{c} \to \mathfrak{h}$ by the composition $\mathfrak{c} \xrightarrow{\sigma} \mathfrak{b} \to \mathfrak{g} \to \mathfrak{h}$. Observe that this map is independent of $\sigma$ since the difference between any two such sections is a linear map $\mathfrak{c} \to \mathfrak{r} \oplus \mathfrak{l}$ which is being factored out. Similarly one can check that the map is one-to-one, hence—counting dimensions—it is a vector space isomorphism. It is now routine to verify that this map is in fact a Lie algebra morphism. The key observation is that the obstruction is now a linear map $\beta : \bigwedge^2 \mathfrak{c} \to \mathfrak{a} \cong \mathfrak{r} \oplus \mathfrak{l}$.

(1) Case: $\mathfrak{c}$ simple

If $\mathfrak{c}$ is simple, $\mathfrak{g}$ is semisimple, which implies that $\mathfrak{r}$ is the radical of $\mathfrak{b}$. The splitting of (A.2) then follows from the Levi-Malcev theorem.

(2) Case: $\mathfrak{c}$ one-dimensional

The splitting of (A.2) is equivalent to the splitting of

$$0 \longrightarrow \mathfrak{r} \longrightarrow \mathfrak{b} \longrightarrow \mathfrak{l} \times \mathfrak{c} \longrightarrow 0 \ . \tag{A.4}$$

Since $\mathfrak{r}$ is solvable, we will prove this by induction on its derived length. If $\mathfrak{r}$ is abelian, then the split follows as a result of the fact that

$$H^2(\mathfrak{l} \times \mathfrak{c}, \mathfrak{r}) = \bigoplus_{i=0}^{2} H^i(\mathfrak{l}; \mathbb{R}) \otimes H^{2-i}(\mathfrak{c}; \mathfrak{r}) = 0 \ .$$

We take as our induction hypothesis that the above sequence splits for every solvable algebra of derived length $< n$. We let $\mathfrak{r}$ have derived length $n$ and consider the exact sequence

$$0 \longrightarrow \mathfrak{r}/[\mathfrak{r}, \mathfrak{r}] \longrightarrow \mathfrak{b}/[\mathfrak{r}, \mathfrak{r}] \longrightarrow \mathfrak{l} \times \mathfrak{c} \longrightarrow 0 \ .$$

It splits by the induction hypothesis since $\mathfrak{r}/[\mathfrak{r}, \mathfrak{r}]$ is abelian and has derived length zero. Let $s : \mathfrak{l} \times \mathfrak{c} \to \mathfrak{b}/[\mathfrak{r}, \mathfrak{r}]$ denote the splitting map and let $s(\mathfrak{l} \times \mathfrak{c}) = \mathfrak{p}/[\mathfrak{r}, \mathfrak{r}]$ where the subalgebra $\mathfrak{p} \subset \mathfrak{b}$ is the preimage of $\mathfrak{l} \times \mathfrak{r}$ under the surjection $\mathfrak{b} \to \mathfrak{l} \times \mathfrak{r}$. Now, the exact sequence

$$0 \longrightarrow [\mathfrak{r}, \mathfrak{r}] \longrightarrow \mathfrak{p} \longrightarrow \mathfrak{p}/[\mathfrak{r}, \mathfrak{r}] \longrightarrow 0$$

splits by the induction hypothesis since the derived length of $[\mathfrak{r}, \mathfrak{r}]$ is $n-1$. Let $z : \mathfrak{l} \times \mathfrak{c} \to \mathfrak{p}$ denote the splitting map. Then the composition $s \circ z : \mathfrak{l} \times \mathfrak{c} \to \mathfrak{p}/[\mathfrak{r}, \mathfrak{r}] \to \mathfrak{p} \subset \mathfrak{b}$ is the desired splitting of (A.4). □